\documentclass[manuscript,screen]{acmart}

\AtBeginDocument{%
  \providecommand\BibTeX{{%
    \normalfont B\kern-0.5em{\scshape i\kern-0.25em b}\kern-0.8em\TeX}}}

\definecolor{burgundy}{RGB}{144,0,32}

\setcopyright{acmcopyright}
\copyrightyear{2018}
\acmYear{2018}
\acmDOI{XXXXXXX.XXXXXXX}

\acmConference[Conference acronym 'XX]{Make sure to enter the correct
  conference title from your rights confirmation emai}{June 03--05,
  2018}{Woodstock, NY}
\acmPrice{15.00}
\acmISBN{978-1-4503-XXXX-X/18/06}

\usepackage{titlesec}

\usepackage{tabularx}
\usepackage{graphicx}

\begin{document}

\title[Trusting the Search]{Trusting the Search: Unraveling Human Trust in Health Information from Google and ChatGPT}

\author{Xin Sun\textsuperscript{1,4}, 
Rongjun Ma\textsuperscript{2}, 
Xiaochang Zhao\textsuperscript{1}, 
Zhuying Li\textsuperscript{3}, 
Janne Lindqvist\textsuperscript{2}, 
Abdallah El Ali\textsuperscript{4},
AND Jos A. Bosch\textsuperscript{1} \\ 
{\small \textsuperscript{1}University of Amsterdam, the Netherlands} \\
{\small \textsuperscript{2}Aalto University, Finland} \\
{\small \textsuperscript{3}Southeast University, China} \\
{\small \textsuperscript{4}Centrum Wiskunde \& Informatica (CWI), the Netherlands} \\ 
}

\makeatletter
\def\ps@headings{%
    \def\@evenhead{}
    \def\@oddfoot{\normalfont\hfil\thepage\hfil}
    \def\@evenfoot{\normalfont\hfil\thepage\hfil}
}
\makeatother
\pagestyle{headings} 

\begin{abstract}
People increasingly rely on online sources for health information seeking due to their convenience and timeliness, traditionally using search engines like Google as the primary search agent. Recently, the emergence of generative Artificial Intelligence (AI) has made Large Language Model (LLM) powered conversational agents such as ChatGPT a viable alternative for health information search. However, while trust is crucial for adopting the online health advice, the factors influencing people’s trust judgments in health information provided by LLM-powered conversational agents remain unclear. To address this, we conducted a mixed-methods, within-subjects lab study (N=21) to explore how interactions with different agents (ChatGPT vs. Google) across three health search tasks influence participants’ trust judgments of the search results as well as the search agents themselves. Our key findings showed that: (a) participants’ trust levels in ChatGPT were significantly higher than Google in the context of health information seeking; (b) there is a significant correlation between trust in health-related information and trust in the search agent, however only for Google; (c) the type of search tasks did not affect participants’ perceived trust;  and (d) participants’ prior knowledge, the style of information presentation, and the interactive manner of using search agents were key determinants of trust in the health-related information. Our study taps into differences in trust perceptions when using traditional search engines compared to LLM-powered conversational agents. We highlight the potential role LLMs play in health-related information-seeking contexts, where they excel as stepping stones for further search. We contribute key factors and considerations for ensuring effective and reliable personal health information seeking in the age of generative AI. 
\end{abstract}



\ccsdesc[500]{Human-centered computing}
\ccsdesc[300]{Human computer interaction (HCI)}
\ccsdesc{Empirical studies in HCI}

\keywords{information searching, generative AI, search engine, human trust perception}

\maketitle



\section{Introduction}

As online sources offer convenient and quick responses, there has been a growing demand for such sources to deliver reliable and user-friendly health-related information. As results, various online information search agents \cite{Cline, Kibirige, mayo} have emerged, becoming important ways for people to obtain health-related information on a daily basis. These search agents assist individuals in acquiring specific information about diseases and treatments, as well as providing general health and well-being advice \cite{SILLENCE20071853, Cline}. This includes guidance on how to maintain a healthy diet, improve physical condition through daily exercise, and tips for a healthier lifestyle. People's trust in online health information does not only affect how they evaluate and perceive such information, but also influences their follow-up decisions to adopt online health advice \cite{SILLENCE20071853,Trust_and_Medical_AI}. Therefore, it is particularly important to explore people's attitudes and trust towards different online search agents and the factors that impact their trust, especially in the sensitive area of personal health.

Google has been a major focus in previous research as it is the dominant search agent in the market \cite{t2} and is thus the main approach where most people obtain online health information. One of research \cite{22} explored users’ behaviors when using Google to search for health information, while the others have delved into users’ attitudes and perceptions during their health information search \cite{6,21}. 
Within the HCI research related to users' perceptions, the trust in Google as the information source and the degree of reliance on specific health information obtained from Google have become the focal points. 
Earlier works \cite{Singal,Sillence4,Sillence,Guo2} have specifically built models targeting trust construction, revealing how various factors individually and collectively influence trust in online health information. In addition to general trust models, other research \cite{Vega,Sillence1,chin2002patients} narrowed its focus on search engines like Google and specifically investigated factors that might influence users' trust in health information retrieved from Google, such as the sources of information. These research offers valuable insights for enhancing human trust in health information seeking scenarios.

With the emergence of generative AI technologies, large language models (LLMs) have gained attention for their excellent performance in dialogue and question-answering tasks \cite{NEURIPS2020_1457c0d6}. As a representative of LLM-powered conversational agents, ChatGPT is widely applied in context of health \cite{t6, t7, t8, t9, t11}.
Despite offering efficient and easy search for health information, these LLM-powered conversational agents also provide individuals with more engaging and user-friendly interactions. These interactions provided by ChatGPT transformed the process of information searching into conversational exchanges with both challenges and opportunities~\cite{Human–AI-Collaboration-in-Healthcare}. The goal of searches has thus evolved from merely returning a list of relevant web pages to directly answering users’ queries. This humanized interaction greatly enhances the user experience of health information search and creates a more personalized search process compared to traditional search engines. Meanwhile, traditional search engines like Google and Bing are also moving in this direction \cite{bing}. They now provide short text snippets (often referred to as answer boxes or quick answers) at the top of the result pages for certain queries. 
This transformation is particularly important in the field of health information search where the value of retrieved information goes beyond convenience and might cause a significant impact on individuals' health and well-being. 
Thus, the applications and potential of ChatGPT are especially noteworthy. This has encouraged more researchers to explore the value of LLM-powered conversational agents in obtaining health-related information \cite{t6, t7, t8, t9, t11}. 
Therefore, in this paper, we consider ChatGPT not merely as a conversational agent, but as a search agent.

Before the advent of LLMs-powered search agents, the connection between people and online health-related information was mostly established in traditional search engines, where their impact on human searching behaviors and trust perception have been extensively studied \cite{SILLENCE20071853}. 
However, trust perception regarding the personal health information provided by ChatGPT has not yet been investigated. Furthermore, while previous studies have touched upon trust-related topics concerning search engines and online health information \cite{SILLENCE20071853,5,8,9,14,21}, there is limited evidence comparing trust perceptions in health information obtained from time-tested search engines versus cutting-edge LLM-powered conevrsational agents.
With this context in mind, our study positions itself at this intersection and proposes the following research questions:
\begin{itemize}
    \item (RQ1) Do people’s perceived trust for personal health-related information differ between traditional search agents (i.e., Google) and LLM-powered conversational agents (i.e., ChatGPT)?
    \item (RQ2) What factors contribute to the perceived trust in personal health information across these different search agents?
\end{itemize}

To unpack these research questions in depth, we conducted a study with a mixed-methods, within-subjects design, including a lab study and semi-structured interviews. During the lab study, participants interacted with two distinct search agents: the search engine (Google) and the LLM-powered conversational agent (ChatGPT) to complete given search tasks in the domain of health. 
The lab study was designed to capture the nuances of participants' interactive behaviors and emergent trust patterns during the search. Subsequently, the semi-structured interviews can enrich our findings by offering a qualitative lens to delve into the complexities of individual experiences and trust perceptions. 
The results showed significant differences in trust perception in the health information, with participants’ trust levels in ChatGPT higher than Google.
More importantly, we identified potential factors influencing people's trust in health-related information from various search agents through the semi-structured interviews, including participants’ prior knowledge, the style of information presentation, and the interactive manner of using search agents.
We further discuss the potential role of LLM-powered agents in the health information seeking context (e.g., ChatGPT as stepping stone for search) and the insights and concerns around machine intelligence and anthropomorphism introduced by LLM-powered search in the context of health information seeking. This provides a new perspective by which to further understand search attitudes and behavior as we increasingly rely on LLM-powered search agents.

The motivation behind this work is to enhance our understanding of human trust perception in health-related information from both traditional search engines and emerging LLM-powered search agents, which is crucial for developing more trustworthy, reliable, and user-friendly search interfaces. 
Our exploratory work makes two key contributions to the community of human-agent interaction and user interface for individual healthcare:
\begin{enumerate}
    \item we examine and differentiate human perceived trust in personal health information acquired from traditional search engines (i.e., Google) versus generative LLM-powered agents (i.e., ChatGPT).
    \item by exploring factors that influence the perceived trust in online health information, we provide empirical insights that can benefit the future development and refinement of LLM-powered search interfaces and agents for health information seeking. 
\end{enumerate}



\section{Related work}

\subsection{Trust in web-based health information seeking}
Online health information searching has become a main way for people to get health advice due to its convenience and speed \cite{Cline, Choudhury, Kim}, attracting a continued increase of using such online platforms for health information. Over the past decades, web-based health information search usually relies on search engines (like Google) and some medical websites \cite{Kibirige, nih, mayo}. Search engines, with their powerful indexing and retrieval capabilities, can offer users a wide range of health information. People can quickly obtain relevant health information by simply entering related keywords, which is why search engines like Google have become the primary tool for most people to retrieve information \cite{Cline,tt}, including health-related information. Additionally, some professional online medical websites, like NIH \cite{nih} and Mayo Clinic \cite{mayo}, aim to provide authoritative and reliable health information and advice, serving as sources of health information online.

The rise of these online platforms, websites and sophisticated search tools has led to people increasingly relying on search agents for health advice seeking. This trend has sparked amounts of research on how people perceive and trust different online health information and its sources \cite{4,5,8,14,22,Vega,Sillence1,chin2002patients}. However, the trust that individuals place in online health information is shaped by a complex array of factors, each contributing to the overall perception of trust. Many existing studies emphasize that the reliability of the information source is very important in this context \cite{Zhang1, Vega, Sillence1, Liu1, Singal, bates2006effect, dutta2003trusted, hesse2005trust,Lucassen}. Information from well-known medical institutions, peer-reviewed scientific journals, or recognized healthcare professionals is more likely to be trusted. On the other hand, websites with user-generated content or commercial motives are often questioned \cite{5,6,7,8,Choudhury}. Alongside source credibility, the intrinsic quality of the information itself also plays a crucial role \cite{2,5,6,8,9,Gil}. Health information that is clear, accurate, and supported by scientific evidence usually earns high trust. Moreover, how the information is presented also affects its trustworthiness\cite{8,9,14,21}. For instance, a well-designed, easy-to-navigate website or app is usually seen as more trustworthy than a platform with a confusing or complicated user interface. Research also shows that users tend to trust search engines' ranking algorithms, often perceiving higher-ranked results as more credible \cite{22,Helena}. However, concerns about the potential for search engine manipulation and dissemination of misinformation have arisen. \cite{t5,Seckler,Haque}. Emerging concerns about data privacy and ethical considerations are also increasingly influencing trust levels \cite{Haque,Friedman,Bansal}. Transparency regarding how personal health data is stored, used, and protected can have a significant impact on human trust \cite{30,Sciascio,Bansal}. 
Furthermore, understanding and measuring people's trust in online information is a tricky task since human perception of trust is complicated to define and measure. Human intrinsic trust is not one-dimensional but is influenced by many factors. Past research \cite{20,Gao1,Liu1,SILLENCE20071853,Sillence3} has carefully investigated how people come to trust in mind, providing valuable insights into the way trust is built. Also, the work \cite{Singal,Sillence4,Sillence,Guo2} explored how the trust model is constructed. This body of work has been crucial in helping us understand, in a more systematic and scientific way, how people can develop trust in online information and the search agents for health information seeking.


\subsection{ChatGPT in health and trust of ChatGPT}

Conversational information seeking \cite{Dalton} steps into the field of health, with ChatGPT demonstrating its capability in handling health-related queries, although research in this area is still in its early stages. A recent systematic review \cite{t6} highlighted several potential advantages of using ChatGPT in healthcare settings, including bolstering health literacy, aiding medical research, and facilitating training of healthcare professionals. More specifically, ChatGPT holds promise in revolutionizing front-line medical services by offering automated patient consultations, preliminary diagnoses, and general health recommendations \cite{t7,t8}. However, there are also concerns about the application of ChatGPT in healthcare. Ethical issues are raised regarding the potential biases of the information and the protection of sensitive health data \cite{t6, t9}. 

Zhu et al. \cite{39} examined ChatGPT's efficacy in responding to queries about prostate cancer. The results revealed that its performance surpassed other large language models. Nonetheless, the study's focus was limited to the accuracy of the responses. Comparisons between these responses and those from traditional search engines, such as Google, were not conducted. Hristidis and colleagues \cite{t10} compared ChatGPT and Google for queries related to dementia and other cognitive declines. Results showed that ChatGPT was rated as more objective and relevant, while Google offered more currency and reliability. Another study by \cite{t11} found that medical experts considered responses from ChatGPT as trustworthy, with 40 percent rating ChatGPT's responses as more valuable than Google. However, all responses from the above studies were assessed by medical service providers and researchers. It is not yet clear whether the general public without much prior knowledge would perceive and evaluate the health-related information differently. Furthermore, most previous studies focused on the quality of the information obtained ChatGPT and Google. Therefore, there remains a pressing need to compare how users interact with and trust both ChatGPT and Google during real-world information-seeking tasks. 

Notably, in comparison to traditional search engines, ChatGPT offers a more personalized and interactive search experience. This different interaction experience from ChatGPT might affect users' trust level on the information obtained. On the one hand, the conversational nature of ChatGPT could lead to personalized perceptions, potentially heightening trust. On the flip side, any erroneous or biased response from such chatbots could significantly undermine perceived reliability, given that users expect consistency in interactions. 

Unlike the extensive literature on Google, the factors influencing people's trust on ChatGPT itself and its responses remain largely unexplored. Studies on other conversational agents such as chatbots suggest that trust is influenced by a myriad of factors. Factors such as the communication style of chatbots, the depth of medical knowledge displayed, and even their "personality" can all substantially impact user trust \cite{t14}. 
Studies suggest that the trustworthiness of ChatGPT may be influenced by its consistency in offering health information \cite{t12} and its transparency \cite{t13}. 

In conclusion, while ChatGPT has showcased immense potential in the health sector, we still need to conduct further research into the level of trust people place in the information provided by ChatGPT, as well as investigate the differing perceptions and trust levels when people use ChatGPT versus traditional search engines, such as Google, for health information searching.


\section{Study methods}

We conducted a mixed-methods study with a lab study followed by semi-structured interviews. For the lab study, we employed a within-subjects design to explore how people interact with two different search agents. Participants were asked to use the search engine Google and LLM-based agent ChatGPT to complete search tasks in the health context. The mixed-methods study combines quantitative and qualitative data, allowing researchers to observe interactions during searching tasks and explore the underlying meaning of trust through interviewing. 


\subsection{Procedure}

Before the lab session, all participants received instructions about the study and informed consent was obtained following our institute's guidelines. 
We employed a within-subjects design with two study conditions: use Google to search and use ChatGPT to search. During the lab study, each participant was asked to complete three search tasks with one search agent, followed by another three search tasks with the other agent. Three search tasks for each search agent are different and pre-selected from each category of search tasks sets (See 3.1.2 for search tasks dataset). The order of agent usage is counterbalanced to reduce the order bias. 
For each search task, participants can freely interact with the search agent until they find a satisfying answer. Participants were allowed to iterate on search queries and ask follow-up questions.
Upon the completion of each search task, we asked participants to rate their trust in the information obtained through a short survey. After completing all search tasks with each search agent (Google or ChatGPT), participants were asked to rate their trust in the search agent itself.

Following the study, we conducted a semi-structured interview. Through understanding participants' interaction experiences with two agents during the study and rationales for rating the trust scores, we explored the underlying factors associated with the dynamic of trust. Each semi-structured interview lasted around 25 minutes. We started the interview by asking general questions about how participants use different tools in their daily lives to search for health-related information. Then we proceed to questions specific to the study interactions with each agent, such as ``How do you validate the information you get from Google?'' and ``How do you feel about the answers retrieved from ChatGPT (sub-topics: satisfied, trustworthy) ?''. At the end of the interview, we discussed the two agents by comparing how the experiences of using them differ and briefly explored potential improvement for the tools. 

\begin{figure}[htbp]
\centering
\includegraphics[width=0.76\textwidth]{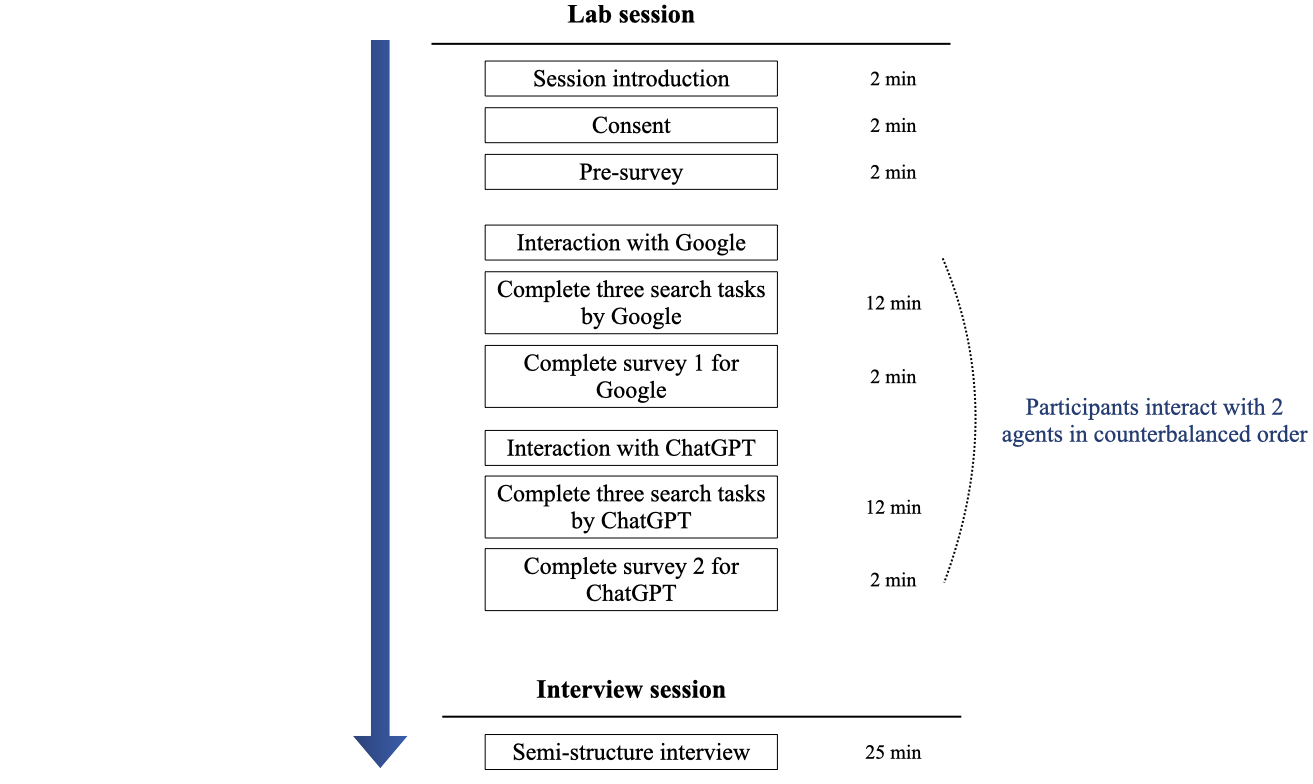}
\caption{The procedure of the lab study and semi-structured interviews.}
\vspace{-2mm}
\label{fig:procedure}
\end{figure}

The overview of the study procedure is outlined in Fig \ref{fig:procedure}. 

\begin{figure}[htbp]
\centering
\includegraphics[width=0.95\textwidth]{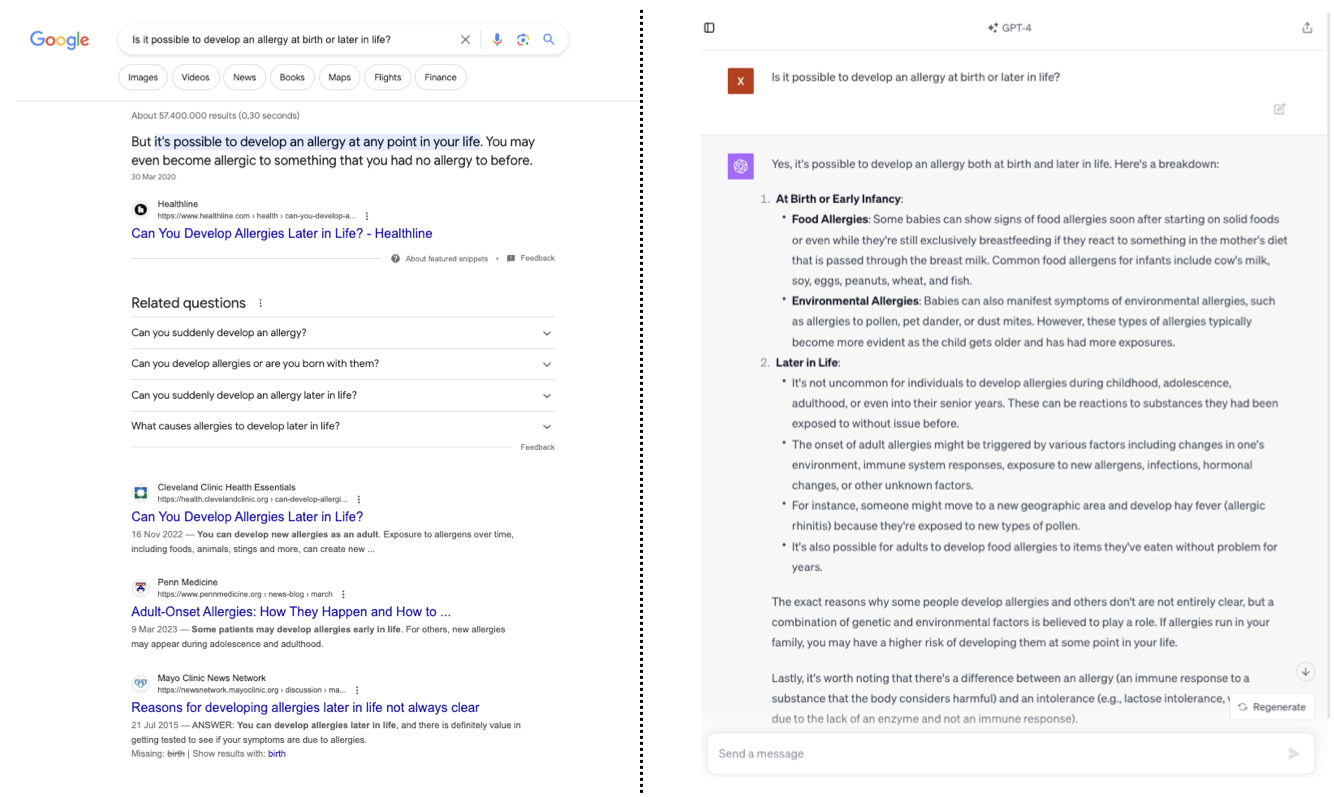}
\caption{Interface of search agents with a search task example using the first type of search task (informational health questions): 'Is it possible to develop an allergy at birth or later in life?'}
\vspace{-3mm}
\label{fig:interface}
\end{figure}


\subsubsection{Search agents.}

For this study, we used two search agents where participants search for health information: Google and ChatGPT (see Fig \ref{fig:interface}). The study was conducted in the university lab in a quiet room with desktop computers running the Windows operating system. Participants accessed both Google and ChatGPT through the Chrome browser (version 117.0.5938.92) by visiting their official websites. For interactions with ChatGPT, participants used the GPT-4 model (July 19 2023 version, without web browsing capability). This setup ensured all participants had a consistent experience while using up-to-date technology for each search agent.


\subsubsection{Search tasks.}
In this lab study, search tasks are open-ended health-related questions that we ask participants to find answers to. 
Each participant was asked to conduct a total of six search tasks during the lab session. To ensure diversity and comprehensiveness, we selected all health-related questions from a large open-sourced dataset \cite{yahoo,CHQ_Summ} of individual health-related questions from the Yahoo forum. Questions in this dataset are all labeled in different categories and we adopted three categories of questions for this lab study based on pre-selection. 50 questions were selected from each category (the list of questions used as search tasks in the study are included as supporting material.) Three categories of questions are described as follows:  
\vspace{-2mm}
\paragraph{Informational health questions.}
These questions aim to gather general knowledge or facts about a specific health topic or problem. Example questions in this category are: "Is it possible to develop an allergy at birth or later in life?"; "What diseases are airborne?"
\vspace{-2mm}
\paragraph{Treatment and advice-related health questions.}
Questions under this category focus on understanding potential treatments or health advice for specific conditions or problems. Two example questions are: "What are the treatments for curing dry eye syndrome?"; "How can I lower my heart rate?"
\vspace{-2mm}
\paragraph{Symptoms and diagnosis-related health questions.}
This group of questions revolves around understanding symptoms associated with particular health conditions and their subsequent diagnoses. Two example questions are: "What are the symptoms of the eating disorder?"; "Can Anxiety make you feel short of breath all the time?"


\subsection{Participants}
A power analysis conducted via G*Power \cite{gpower} indicated that a minimum of 20 participants would be necessary to detect a medium effect with an alpha level of .05 and a power of 80\%. We recruited 21 participants (N=21) through the on-campus recruitment system at the institute. Participation was entirely voluntary. Informed consent was obtained from each participant prior to the lab session. Monetary compensation (10 EUR) was provided to participants in accordance with institute guidelines. The inclusion criteria specified that participants must be fluent in English and have experience with online information searching. The study was approved by our institute's ethics and data protection committee.


\subsection{Measures}
Before the lab study, we inquired about basic information from participants, including demographic information and their usage experience on different search agents. During the study, We measured the level of trust by participants in the health-related information acquired from the two search agents with short surveys. Here, we illustrate measurements to collect the information mentioned above.  
\subsubsection{Before lab study}

\paragraph{Demographics and prior experience.} 
\vspace{-2mm}
Before the lab study, we collected basic demographic information of participants such as age, gender, education, and field of study or occupation. Additionally, we also asked how participants use online resources for health information, example questions are 'What search agent(s) do you use primarily for searching information online?' and 'How often do you use the Internet as a source of health information?'.

\paragraph{Propensity of trust in technology.}
To understand participants' inherent disposition towards trusting technology, we surveyed participants with a pre-study questionnaire. This questionnaire measured individuals' general tendency to rely on and have confidence in technology across various contexts. The propensity of trust in technology scale adopted from \cite{ppt} consists of 6 items measured on a 5-point Likert scale ranging from 1 (Strongly Disagree) to 5 (Strongly Agree). An example item is: "I think it’s a good idea to rely on technology for help."

\subsubsection{Upon completing each search task}
\paragraph{Human perceived trust in health-related information.}
\vspace{-2mm}
After completing each search task, participants were required to rate their trust level in the health-related information acquired from the respective search agent. The validated questionnaire ('Trust of online health information' \cite{ti,Rowley2015StudentsTJ} was adopted to evaluate the perceived credibility, reliability, and believability of the information content itself, independent of its source or the search agent. Eleven items measured on a 5-point Likert scale ranging from 1 (Strongly Disagree) to 5 (Strongly Agree) were employed for this purpose. The cronbach's alpha of this scale in our study is .95. One example item is: "The information appears to be objective."

\subsubsection{After each study condition}

\paragraph{Human perceived trust in the search agent.}
\vspace{-2mm}
In addition to measuring trust in information, we measured participants' trust in the search agents as well to investigate the relationship between the trust in information and the trust in the search agent itself.
After completing all three search tasks with each search agent, participants were asked to rate their trust in the search agent's ability to deliver health information. We adopted the 'Model of online trust of health information websites' questionnaire \cite{tt} with 15 items using a 5-point Likert scale ranging from 1 (Strongly Disagree) to 5 (Strongly Agree). This measure aimed to capture the broader trust towards the agent, encompassing aspects like usability, satisfaction, and overall reliability. The cronbach's alpha of this scale in our study is .62. One example item is: "The search agent (e.g., Google) provides truthful information."

\paragraph{Intention to use search agent for health-related information searching.}
We used a self-constructed questionnaire with one item to measure the participant's intention to use the specific search agent for health-related information searching with a 5-point Likert scale ranging from 1 (Strongly Disagree) to 5 (Strongly Agree). The item is: "I would intend to use Google (or ChatGPT) for health-related information seeking."


\subsection{Data analysis}
We collected a mix of quantitative and qualitative data in this study to get an overview of how people interact with two different search agents for health information seeking and whether there are differences in the trust in the health information they obtained from the two search agents.

Starting with the quantitative data from the survey, we checked to ensure our data met certain statistical requirements, which validated the robustness of any subsequent statistical tests and interpretations. We did Shapiro-Wilk test \cite{SHAPIRO1965} to check the assumption of normality 
and Bartlett's test \cite{Arsham2011} to check the assumption of homogeneity of variance. Both assumptions were not violated.

After the descriptive analysis summarizing the main points of the data, we applied a two-way repeated-measures ANOVA analysis, and a paired sample t-test to find out any significant differences in trust perceptions when participants interacted with two distinct search agents. Besides, the correlation analysis was conducted as well to explore any correlation among the variables.

For the qualitative interview data, we conducted thematic analysis \cite{braun2012thematic} to explore the factors influencing people's trust in online health information retrieved from different search agents. As a preliminary step to the coding process, the first three authors transcribed the interview recordings, totaling 516 minutes, involving 21 participants. To develop a codebook, the first three authors initially open-coded two identical transcripts and discussed them to reach an agreement on the codes. During this step, similar codes were merged, and some code names were refined through consensus. Following the agreed-upon codebook, the three authors coded the remaining interviews. Throughout the analysis, they discussed and peer-reviewed each other's coding processes. Based on the initial codes, the authors collectively discussed and collated the codes into potential themes. Through iterations of reviewing and refining the themes, they defined and named them collaboratively. Through iteration of reviewing and refining themes, together they defined and named the themes. In the following findings section, we will report identified themes in response to research questions.


\section{Findings}


\subsection{Lab study}
\vspace{-2mm}


\begin{table}[!h]
\centering
\renewcommand{\arraystretch}{0.82}
\begin{tabularx}{\columnwidth}{p{5cm} >{\raggedright\arraybackslash}p{6cm} >{\raggedright\arraybackslash}X}
\toprule
\textbf{Demographic} & \textbf{Categories} & \textbf{Numbers of Participants (\%)} \\
\midrule
Gender & & 
\\ 
 & Female & 16 (76.1\%)
\\ 
 & Male & 5 (23.9\%)
\\
\hline
Age & & 
\\
 & 18-24 & 15 (71.4\%)
\\
 & 25-34 & 5 (23.8\%)
\\
 & 65+ & 1 (4.8\%)
\\
\hline
Education & & 
\\
 & High school & 1 (4.7\%)
\\
 & Bachelor & 9 (42.9\%)
\\
 & Master & 10 (47.6\%)
\\
 & Doctor & 1 (4.7\%)
\\
\hline
Professional Domain & & 
\\ 
 & Social Science & 10 (47.6\%)
\\
 & Business and Commerce & 4 (19.0\%)
\\
 & Health and Medical Science & 1 (4.7\%)
\\
 & Computer Science \& Information Technology & 3 (14.3\%)
\\
 & Other & 3 (14.3\%)
\\
\hline
Frequency of online health \\information seeking & & 
\\
 & Often & 9 (42.8\%)
\\
 & Sometimes & 11 (52.3\%)
\\
 & Rarely & 1 (4.7\%)
\\
\hline
Frequently used search agent & & 
\\
 & Search engine & 21 (100\%)
\\
 & Conversational agents & 15 (71.4\%)
\\
 & Social media platforms & 8 (38.0\%)
\\
\bottomrule
\end{tabularx}
\vspace{0.3mm}
\caption{Characteristics of participants}
\vspace{-6mm}
\label{table:participants}
\end{table}


\subsubsection{}{Participants characteristics.}
We recruited 21 participants through the institutional recruitment platform. Our participants range from 18 to 65 years old, with the majority age group of 18-24 (71.4\%). 
Among the participants, approximately 47\% hold undergraduate degrees, the remaining 47.6\% have attained postgraduate qualifications and 4.7\% have a doctorate degree. 
Participants are from diverse professional domains, 47.6\% come from social science, 19\% from business and commerce, 14.3\% from computer and information technology, and the remaining 14.3\% represent a mix of other professions. 42.8\% of our participants frequently turned to online sources for health-related advice, and 52.3\% sometimes searched for health information online. In contrast, only 4.7\% reported that they rarely or never rely on online resources for health information. 
Characteristics of participants are summarized in Table \ref{table:participants}.


\subsubsection{Descriptive statistics.}
We measured the propensity of trust in technology, trust in health information from two agents, and trust in the search agents. We ran a descriptive data analysis on these measurements, Table \ref{table:descriptive} presents an overview of the results.

The mean values for each variable indicate the average level of trust participants reported in each category. The mean score for the propensity of trust in technology (M=3.62, SD=.76) shows our participants have a positive trust in general technology.
We measured participants' trust in retrieved information after each task, in general, participants showed higher trust in information retrieved from ChatGPT. The trust in information retrieved from two search agents is different. The average trust score of the three tasks is 3.77 (SD=.64) for Google and 4.05 (SD=.47) for ChatGPT. Additionally, we measured participants' trust in the search agent after each experiment condition, and participants showed higher trust in ChatGPT. The average trust in search agent Google is 3.31 (SD=.99) and ChatGPT 3.58 (SD=.98). 


\begin{table}[htbp]
\centering
\renewcommand{\arraystretch}{1.25}
\begin{tabularx}{\columnwidth}{p{7cm} >{\raggedright\arraybackslash}X >{\raggedright\arraybackslash}X}
\toprule
\textbf{} & \textbf{Google \newline Mean (SD)} & \textbf{ChatGPT \newline Mean (SD)} \\
\midrule
Propensity of trust in technology & \multicolumn{1}{r}{3.62 (.76)}
\\ 
\hline
Trust in information (search task 1) & 3.69 (.90) & 4.01 (.74)
\\ 
Trust in information (search task 2) & 3.81 (.89) & 4.10 (.74)
\\ 
Trust in information (search task 3) & 3.82 (.88) & 4.03 (.81)
\\ 
Trust in information (Average of three search tasks) & 3.77 (.64) & 4.05 (.47)
\\ 
Trust in search agent & 3.31 (.99) & 3.58 (.98)
\\
Intention to use for health information searching & 3.57 (.49) & 3.52 (.85) 
\\
\bottomrule
\end{tabularx}
\vspace{1mm}
\caption{Descriptive statistics of trust perception in health information and search agent. Search task 1 is Informational health questions, 2 is Treatment and Advice-related questions, 3 is Symptoms and Diagnosis-related questions.}
\vspace{-5mm}
\label{table:descriptive}
\end{table}


\begin{figure}[htbp]
\centering
\includegraphics[width=0.98\textwidth]{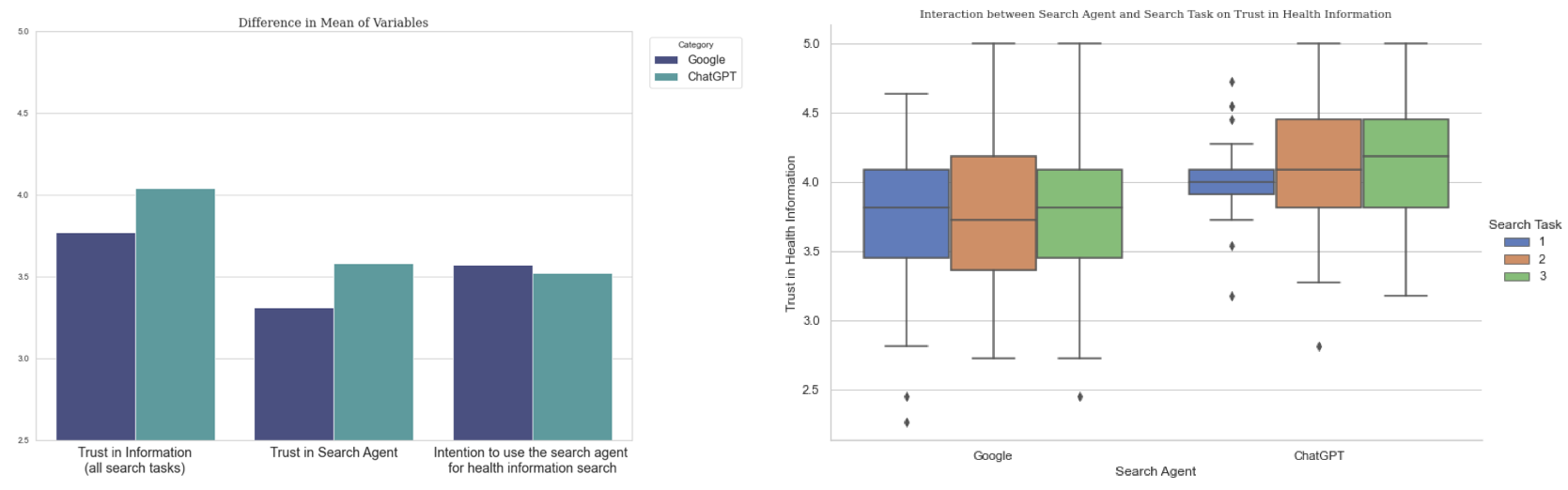}
\caption{Left: Mean of trust in health information, the search agent and the intention to use the search agent for health information search. Right: Mean of trust in health information between two search agents across three search tasks.}
\vspace{-1mm}
\label{fig:mean-difference}
\end{figure}


\subsubsection{Human trust in health-related information differs by search agents.}
We used a two-way repeated-measures ANOVA to compare the difference in trust between the two agents.
The analysis was performed to explore if there are differences in trust, including trust in the retrieved information and search agents.
The objective was to ascertain if the observed differences between the variables were statistically significant or could be attributed to random chance, the results are shown in Table \ref{table:anova}.

The main effect of the search agent on trust in health-related information was found to be significant, with \( F(1, 20) = 6.73, p = .017 \), and \( \eta^2 = .057 \). This indicates a statistically significant difference in trust levels in health information between the two agents: Google and ChatGPT. 
The trust score for ChatGPT is marginally higher compared to Google, 
signaling a user tendency to rely more on ChatGPT’s outputs for health-related advice.
Regarding the main effect of the search tasks, the results were not statistically significant, \( F(2, 40) = 0.63, p = .538 \), with \( \eta^2 = .006 \). This finding implies that the variations in trust levels were not significantly influenced by the different categories of search tasks.
The trust in the retrieved information remained consistent regardless of the category of health-related questions.
The interaction effect between the search agent and search task was not significant, as evidenced by \( F(2, 40) = 0.20, p = .817 \), and \( \eta^2 = .002 \). This non-significant interaction suggests that the difference in trust levels between the two search agents does not vary significantly across different categories of search tasks. Thus, the trust in each search agent remains relatively stable and consistent across various types of search tasks, as shown in Fig \ref{fig:mean-difference}.


\subsubsection{Human trust in search agents differs.}
To explore the human perceived trust in the search agents, we did a paired sample t-test. The trust scores in Google and ChatGPT as search agents are different. The difference is statistically significant with the difference in means as 0.27, $t$(21)=-2.53, $p$=.02, Cohen's $d$=0.55 (see in Fig \ref{fig:mean-difference}). This indicates that the overall trust people have in these two search agents as information-seeking tools is significantly different. 


\begin{table}[htbp]
\centering
\renewcommand{\arraystretch}{1.25}
\begin{tabularx}{\columnwidth}{p{4.5cm} >{\raggedright\arraybackslash}X >{\raggedright\arraybackslash}X
 >{\raggedright\arraybackslash}X >{\raggedright\arraybackslash}X >{\raggedright\arraybackslash}X >{\raggedright\arraybackslash}X >{\raggedright\arraybackslash}X}
\toprule
\textbf{Source} & \textbf{SS} & \textbf{ddof1} & \textbf{ddof2} & \textbf{F} & \textbf{p-value} & \textbf{eps} \\
\midrule
Search Agent & 2.34 & 1 & 20 & 6.73 & \textbf{.017} & 1.00 \\

Search Task & 0.24 & 2 & 40 & 0.63 & .48 & .68 \\

Search Agent \textasteriskcentered{} Search Task & 0.06 & 2 & 40 & 0.20 & .78 & .83 \\
\bottomrule
\end{tabularx}
\vspace{1mm}
\caption{Two-way repeated measures ANOVA results. SS (Sum of Squares): represents the variability in the dependent variable "Trust in Health Information". F (F-statistic): to determine whether the variability between group means is larger than the variability within the groups, higher F-values usually suggest the groups are different. P-value: represents the significance of the F-statistic. eps (Epsilon): Sphericity correction factor.}
\vspace{-6mm}
\label{table:anova}
\end{table}


\subsubsection{Correlation of trust across search agents.}
\begin{figure}[htbp]
\centering
\includegraphics[width=0.96\textwidth]{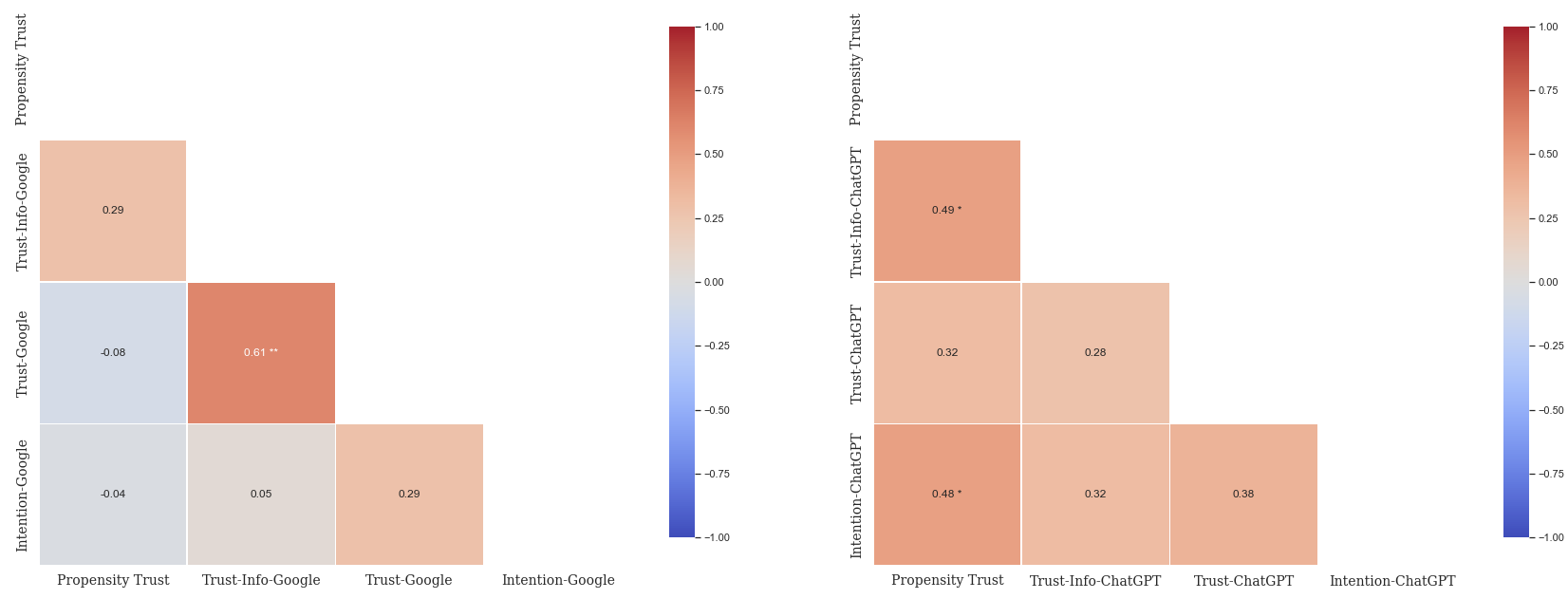}
\caption{Pearson correlation with key variables (Left: Google; Right: ChatGPT. **$p$<.01, *$p$<.05).}
\vspace{-3mm}
\label{fig:correlation}
\end{figure}

We conducted a Pearson correlation analysis to analyze the relationships between the key variables in this study, see Fig \ref{fig:correlation}. 

We found some interesting relationships. First, trust in health information from Google shows a significant positive correlation with trust in Google, $r$(21)=0.61, $p$=.003, suggesting that participants who have higher trust in health information retrieved from Google are also more likely to trust Google as a search agent. Trust in health information from ChatGPT also shows a positive correlation with trust in ChatGPT but not significant, $r$(21)=0.28, $p$=.21.
Second, there is a significant positive correlation between the propensity of trust in technology and trust in the health information from ChatGPT, $r$(21)=0.49, $p$=.02, but no significant correlation between the propensity of trust in technology and the trust in search agent ChatGPT, $r$(21)=0.32, $p$=.16. This implies that individuals with a general tendency to trust technology also have higher levels of trust in the health information retrieved from ChatGPT. However, there is no significant correlation between the propensity of trust in technology and trust in the information from Google, $r$(21)=0.29, $p$=.20, and search agent Google, $r$(21)=-0.08, $p$=.73.
Third, we observe that the intention to use ChatGPT for health information searching is positively correlated to the propensity of trust in technology, $r$(21)=0.48, $p$=.026, however, Google does not have such correlation, $r$(21)=-0.04, $p$=.87. 

The correlation analysis not only strengthens our understanding of how different facets of trust are interconnected, but also lays the groundwork for more complex analyses and interpretations in our qualitative findings. 


\subsection{Semi-structured interviews}
In this section, we report themes generated based on interview data. Themes are organized under three sections: \textit{How do people search for health-related information} gives an overview of real-life scenarios, showing the purpose of people searching online and individual differences in sourcing the answers. \textit{What factors influence trust in information} illustrates diverse factors that shape the trust in information online and compares how the difference between two agents affects people's trust. Lastly, \textit{How do people want ChatGPT to improve} points to the potential improvement for ChatGPT to better support information searching.

\subsubsection{How do people search for health-related information?}
\paragraph{Searching online is a pre-step to deal with health-related problems.} Our participants search for different health-related topics online in their daily lives, but none of them have major concerns regarding health conditions. The four main types of health-related topics for online searching are lifestyle questions such as \textit{diet and exercise} (P2, P14-15, P17, P19, P21), accident treatment such as \textit{tore a muscle} (P2), personal health conditions such as \textit{allergies} (P6), and mental health problems (P4). As the medical problem is a serious area for most participants, they are cautious about the information-searching process. People consciously hold back a degree of trust toward online health information:
\begin{quote}
    \textit{I always make sure to keep it to keep in mind that. Umm, it's not a reliable source. Everything I see on the Internet isn't necessarily true. - P4}
\end{quote}
Usually, people would not expect a perfect answer to directly solve the problem; however, the purpose of searching is to remind themselves of some knowledge they already know, or study the general knowledge on the topic and get a direction for the next step. For example, one participant is hesitant when referring to the medical specialist, and searching online helps direct her through the process doctor consultation.
\begin{quote}
    \textit{
    Google always gives you like an overview of where you should go most of the times, they do recommend the space. Yeah. Well, you're having like searching symptoms. – P9}
\end{quote}

Subsequent to the information-searching process, participants selectively adopt online advice 
depending on the importance and urgency of health matters, as well as the practicality and risks of the advice:
\begin{quote}
    \textit{So if it says like it take a pill or something like that, I won't take it. But if says like, drink water, stay in bed, of course I'll do that. - P10}
\end{quote}

\paragraph{People use their own combinations of sources for health-related information.} 

People have their own preferences for information sources regarding health-related topics, our participants reported diverse information sources such as forums, social media, and websites. With their own experience in searching for health-related information, people acknowledge the high credibility and practicality of specific resources. For example, many participants share high trust in official web domains such as the NHS website, and Google Scholar (P1-2, P4-12, P14-21). Some participants naturally trust big company brands such as Google (P7). There are also participants who use social media as a reliable source of medical information because it allows them to match symptoms more easily (P11, P17, P19):
\begin{quote}
    \textit{There might be some, uh, like influencers. They are from this area, they just share health-related information. [...] Sometimes you have some certain pain that it's hard to describe, but people might experience the same and then you can look for what happened to them. - P11 }
\end{quote}
Additionally, people do not rely on just a single resource but a combination of multiple sources. Comparing the results from multiple sources helps our participants validate the information:
\begin{quote}
    \textit{I think sometimes you might need to read a few sources just to be sure. You don't wanna just read one website, you wanna just like double-check if another says the same, make a comparison. - P9}
\end{quote}

\subsubsection{What factors influence trust in information?}
\paragraph{Prior knowledge influences people's trust in information.} 
Although people search for health topics through diverse information agents, the judgment on whether retrieved information is trustful or not largely depends on their prior knowledge. The prior knowledge can be accumulated through people's domain knowledge on the searched topics, as well as previous experiences of using different agents. Previous negative and positive experiences are associated with people's current trust in search agents. A positive experience of recovering following online advice boosts trust, in contrast, the negative experience of misinformation leads to the uncertainty of information: 
\begin{quote}
    \textit{chatGPT can give me some results really quickly. But if I just copied the DOI number and double-checked it, you'll find they're totally wrong. Everyone, they're totally wrong.[...]but those results they pretend to be quite reliable. [...] So that's one of the important reasons I don't fully trust ChatGPT. - P8}
\end{quote}
Compared to Google, ChatGPT is a new tool that most participants do not yet have much experience of usage. While people are still exploring how to search for information with ChatGPT, their trust can be influenced by news and rumors (P21). On the other side, though many participants use Google as a default tool, the experience of too much commercial information (P1-2, P4, P10-11, P13, P16, P18-19, P21), and filter bubbles (P7) \cite{Filterbu39:online} that information is isolated in order to personalize searches also shape people's trust in information from Google.  

\paragraph{Information presentation influences people's trust in information.}
Our participants think health is a serious domain, and whether the information is presented in a professional but understandable form is an essential factor to trust. The presentation of information includes forms of language expression, structure, and visuals.

First, the medical field is professional, and terminologies are often introduced in explaining health information. Our participants think professional usage of language with terminology increases the credibility of information, however, such terminologies also create difficulty for participants to understand the meaning of the searched topics. In addition to professional terminology, the tone of language expression also shapes people's trust, including confidence in expression and passive voice usage. The passive voice of language expression helps build the objectiveness of information and thus leads to higher trust. Furthermore, the confidence of tone is especially mentioned regarding interaction with ChatGPT. As ChatGPT interacts in a more human-like way, how confidently it delivers the information influences our participants' trust. An uncertain answer including \textit{maybe} or \textit{I'm not sure} (P10) will lower the trust, however, a too-certain answer is neither trustworthy 
\begin{quote}
   \textit{if it's too, too confident in stating this and this leads to this [...], (I'll) trust less, Because the answer it's never so straightforward. - P16} 
\end{quote}
Second, the structure of information is a commonly mentioned topic when referring to the interaction with ChatGPT. Whether the information is consistent and logically connected is an important criterion to evaluate the information structure. A well-structured answer guides participants to understand the searched question and increases the information reliability:
\begin{quote}
    \textit{the answer is quite like structured, like it starts from explaining the disease and then give you the symptoms. And it's also like almost list out all the answers, like the major answers, and it will also add in the end like there might be more. Yeah, I think the way chatGPT gave you the answers make you feel it's more reliable. - P11}
\end{quote}
Third, visual elements also play a role in trust. People trust professions in the health domain, and visual elements influence our participants' perceptions of the professional level of information sources. For example, \textit{very colorful} (P14) and {unprofessional logos of the websites} (P4) lead to lower trust regardless of the content of information. 

\paragraph{Interaction influences people's trust in information.}
The major difference between searching with Google and ChatGPT happens in interaction. Our participants shared experiences that searching with ChatGPT is like interacting with humans. ChatGPT understands questions well and gives straightforward answers with a proper amount of details on the background information. 
\begin{quote}
   \textit{
   it's super easy like I can write out a whole paragraph describing, you know, like, oh, my foot hurts a little bit, but not that much, but like when I bend it like this, it hurts. And Google would never understand that.But yeah, you can just you can just spew all of this into ChatGPT and it will understand what it is.  - P2} 
\end{quote}
Furthermore, the follow-up questions are also context-aware and guide our participants well in understanding information. In comparison, searching with Google gives our participants autonomy in selecting information, however, requires efforts in summarizing and synthesizing in order to get an answer. Such difference results in a generally better experience of interaction with ChatGPT and also influences our participants' understanding and trust in the information. In most cases, control of the search process increases people's certainty and trust in the answer. Still, our participants shared that if they have enough domain knowledge to evaluate answers from ChatGPT, a straightforward answer from ChatGPT is preferred and they will not take further steps to verify the answer. 

Usually, the trust with human like conversations also depends on the topic:
\begin{quote}
    \textit{I think that depends on situation. Sometimes I might prefer the human like conversation, but I think there's also depends on what I want to know. If like if I ask a mathematical formula and then ChatGPT used a very human like way to tell me I my feel is not very professional. - P7}
\end{quote}

\subsubsection{How do people want AI to improve?}
\paragraph{Add features to support information searching and understanding.} A major limitation shared by many participants of ChatGPT in terms of trust is the source information. The lack of reference or even faking the references results in less trust in the information. Thus, a main feature proposed by our participants is to add references and redirect people with links to sources, so they can verify information. One potential scenario is to use ChatGPT as a pre-step before searching with Google:
\begin{quote}
    \textit{But also if I didn't know anything about a health issue, I could look it up on the ChatGPT first to also like get a sense of direction. What this could be related to and then continue my search on Google. - P1}
\end{quote}
Additionally, supplementing texts with more rich information including pictures or videos can also be helpful to understand the information, especially in the context of health. For example, using pictures to show \textit{where is the pain} (P5). 

\paragraph{Tailor searches to regional and individual differences.}
Health information diverges in regions such as differences in medical systems in different countries (P2), and individuals such as personal symptoms or allergies (P14). In the context of health, our participants wish ChatGPT to tailor answers with regional and individual differences into consideration. For example, customize the training with regional data (P2, P20), work as a mediator to coordinate the appointments with doctors, and proactively ask about individual situations like human doctors (P2, P14):
\begin{quote}
    \textit{You tell the ChatGPT or tell the human doctor your symptoms and the doctor will ask you some relevant questions about your symptoms and maybe then, it can also find more specific information for you. Maybe it's a different story if you're a man and you're 80 years old, then if you're a woman and you're 18 years old like that, maybe it would give you different information. - P14}
\end{quote}


\section{Discussion}

\subsection{Differential trust levels: ChatGPT vs. Google}
The key findings of our study answered the first research question we posed: there are notable differences in how people trust health information from different search agents. Our participants trusted the health information provided by ChatGPT more than the information given by Google. This trend might suggest that as AI technology improves, people might lean more towards using these new tools and placing trust in the information they provide.
A reasonable explanation for why ChatGPT is trusted more could be its quick response and unique conversational interface \cite{Rheu}. ChatGPT can rapidly adjust to users’ questions, which might be seen as a sign of expertise and relevance. Also, the conversational interaction of ChatGPT may give people a feeling of personalized attention and understanding, making them feel heard and connected \cite{Zhongxuan}. 
The appearance of the user interface also plays a role in building trust \cite{21}. An intuitive and user-friendly interface is easy to use and thus increases its credibility \cite{8}. Furthermore, with AI technologies advancing rapidly, people might link AI-driven tools like ChatGPT with cutting-edge information, influencing their trust \cite{HALEEM2022100089}.
Interestingly, there is no significant difference in trust in the information for different types of health search tasks. This observation suggests that people’s trust may not be tightly connected to the details of specific search tasks. Instead, it might depend on their general trust on the agents or tools they use. 
Therefore, efforts should be made to enhance the perceived reliability and credibility of the agents or technology overall.

Furthermore, we found a significant correlation between trust in health information provided by Google and trust in Google itself. This result might indicate a sign of traditional or longstanding trust in Google. Over the years, people have developed a good relationship with Google, leading to confidence in both the agent and the information it provides \cite{Schultheis2023-va,Haque}. In contrast, there is no such significant correlation for ChatGPT, suggesting that people might separate their trust on ChatGPT from their trust in the information it provides. This could mean that even though ChatGPT has advanced features, people are still taking time be develop a stable trust relationship with the new searching approach.
Interestingly, there is a clear correlation between individuals' inherent trust in technology and their trust in the health information from ChatGPT. This highlights a broader social trend: as the number of digital natives (including the young and well-educated participants in our study) increases, search agents using new technology might naturally benefit from the inherent trust in technology \cite{KESHARWANI2020103170}. 
Finally, there is a positive correlation between individual intention to use ChatGPT for health information seeking and their trust in the health information from ChatGPT, especially among those with a higher inherent trust in technology. This reflects that users who generally trust technology are more likely to consider, accept, and trust LLM-powered technology for their health information seeking. 

Understanding the reasons behind the higher level of trust in ChatGPT is not just important for academic research. It also provides practical insights for designing and developing future LLM-driven health information systems and smart interfaces, by identifying specific features that build trust in people.


\subsection{Anthropomorphic intelligence influences trust in health information}

The anthropomorphic intelligence and features introduced in the interactive search process by LLM-powered search agents (like ChatGPT) have been linked with people perceived level of trust \cite{Pelau, PELAU2021106855, HALEEM2022100089, t11, t7, Korteling}. Therefore, we discuss the potential role and concerns of anthropomorphise in ChatGPT within the health-information searching context.

One obvious anthropomorphic feature of ChatGPT is its human-like language style. Talking in a human-like manner enhances interactive communication and gives users the feeling of “consulting a doctor” or “talking to a friend”. Such a communication style makes information delivery more intuitive and easier to understand, creating a trusting atmosphere between human and technology. 
The role-playing capability is another unique feature of ChatGPT that could enhance its human-likeness in the language \cite{roleplay-2}. Through different prompts, LLM-powered agents can exhibit various performance characteristics, providing a novel approach brought by LLM technology. For instance in \cite{roleplay-1}, when ChatGPT takes on the role of a doctor, it might improve the performance of LLMs by giving more relevant health-related information, thereby enhancing the perception of trust. Such capability of human-likeness in ChatGPT is noteworthy when we investigate the effect of its anthropomorphise.
Besides, as shown in the interview findings, how confidently ChatGPT delivers the information influences participants' trust. ChatGPT can both apologize for and insist on its incorrect response. The balance between being submissive and assertive in the human-like language should be carefully considered as well. However, the appropriateness of human-like language depends on the situation, as mentioned by participant “\textit{I think that depends on situation. Sometimes I might prefer the human like conversation [...] (P7)}”. 

Furthermore, ChatGPT can precisely cater for individual needs which go beyond merely matching keywords and phrases. The ability to directly identify questions from long or ambiguous paragraphs allows for a closer alignment with human needs compared to Google. Such anthropomorphic intelligence grasps the real intent behind user queries, as mentioned by our participants, "\textit{it's super easy like I can write out a whole paragraph describing, [...] And Google would never understand that. But yeah, you can just spew all of this into ChatGPT and it will understand what it is. (P2)}". This fosters a sense of understanding and leads to increased trust in the information. 
Further, ChatGPT can respond to people's follow-up questions while considering the context with the advanced capability of LLM \cite{t12}. People may feel that ChatGPT understands their issues instead of not just addressing isolated questions, as expressed by our participants "\textit{[...] ChatGPT will understand what it is. (P2)}". This experience deepens the connection between human and the agent, influencing the trust in information.

In addition, sensitivity to individual situations might also be a crucial factor and potential improvement of ChatGPT in affecting users' trust in the information, especially in the health context. For example, if ChatGPT can provide personalized responses based on people's symptoms or emotions, people might feel the agent understands them better and offers more targeted help. It not only demonstrates AI’s “empathy” but also improves the relevance of information. Participants also mentioned that with health contexts varying by region or country, the current training data for ChatGPT primarily comes from English-speaking countries and thus might provide biased responses. Therefore, giving answers that adapt to regional differences could be incorporated to build trust in the health information provided by ChatGPT. 

While the machine intelligence and anthropomorphise of LLM agents might positively influence human trust in the information retrieved, we must remain cautious, especially in the field of health. Human-like AI might mislead people and spread incorrect information \cite{Johnson, t9}. When people overtrust AI agents, they might blindly accept the advice provided by AI, leading to decisions that are harmful to their health. Additionally, if people perceive AI agents as “individuals” rather than tools, assigning responsibility becomes blurry when such tragic incidents arise. Therefore, potential ethical issues need to be considered while developing the anthropomorphic features of AI \cite{Ethical}.


\subsection{Searching autonomy influences trust in health information}
\subsubsection{Trade-off between autonomy and convenience.}
People need to strike a balance between autonomy and convenience during health information search, both of which differently impact trust perception. Previous research has shown that\cite{ellis1993comparison}  usability (e.g., ease of use) affects the feeling of trust in the search agents. In our interviews, some participants prefer convenience than autonomy during the search, "\textit{But also if I didn't know anything about a health issue, I could look it up on the ChatGPT first [...] and then continue my search on Google. (P1)}". This inclination arises when participants wish to receive direct and quick answers to efficiently obtain necessary health information, especially for general knowledge or non-serious issues. In this aspect, ChatGPT offers a much better experience than Google because Google requires a trial-and-error approach in the searching process. The study by Xu et al. \cite{xu2023chatgpt} revealed that people spend less time on ChatGPT than Google for all search tasks. In such cases, autonomy in the search process might not be the leading factor influencing trust when our participants prioritize efficiency and convenience.
On the other hand, some participants highly value autonomy particularly when it comes to crucial health-related matters. They believe that having more control and understanding of the search process allows for self-verification of the information. In these cases, autonomy not only allows people to navigate and filter information based on their preferences and needs, but also helps them understand the origins of the information more clearly. The greater clarity enhances the transparency and credibility of the information acquired as proved in \cite{t13}. Recognizing this, designers and developers of LLM-powered search agents need to balance human autonomy and convenience for the search delicately.

\subsubsection{Relation between autonomy and the prior knowledge.}
The importance of autonomy in the trust perception of health information could be affected by prior knowledge as well. In our interviews, some participants emphasized that autonomy is particularly important when they lack prior knowledge about the health questions. They believed that having autonomy during the search can enhance their confidence in the information obtained from the search agents, because the process of independent search provides a sense of security that they can actively assess and verify the reliability of the information. A key insight is that autonomy might compensate for a lack of prior knowledge, serving as a tool for enhancing the trust. This is in line with previous research where users demonstrated an overreliance on ChatGPT's inaccurate responses \cite{xu2023chatgpt}. Without prior knowledge of the search topic, this lack of autonomy from ChatGPT resulted in the inability to verify the information provided and eventually trust on inaccurate answers. On the contrary, our participants with prior knowledge of the search questions placed less emphasis on autonomy in the search process. They claimed that the answers provided directly by ChatGPT might be sufficient, with prior knowledge acting as an internal calibrator for perceived trust. They can rely on their prior knowledge to verify and assess the health information from ChatGPT, reducing the need for complete autonomy. 

From these interview findings, we can speculate that there is a close relationship between one's prior knowledge and the need for autonomy in the context of health information searching. For people who prefer to delve deeper into research and verification of information, autonomy might consistently be a core demand. However, for those seeking quick and direct answers, prior knowledge might reduce dependence on autonomy. These subtle dynamics offer a richer understanding of how autonomy operates in different searching scenarios with various individual situations.


\subsection{Scenarios of using LLM-powered search agent: ChatGPT as a stepping stone}

Health is a serious domain, and people are inclined to be cautious about information searching. Although people generally show a positive trust in technology, they reserve trust during health information searching and “\textit{keep in mind [...] Internet isn't necessarily true (P4)}”. Regardless of search agents where the information is from, the validation of information is essential before utilizing it. The validation process of building trust depends on people's prior knowledge, including prior domain knowledge (aligned with ~\cite{domain_search}), as well as a biased baseline from previous experience using the tool. Similar to Ruju and Reilly's \cite{raju1980product} finding that past experience influences an individual's decision-making, the negative experience of receiving fake answers from ChatGPT also leads to less trust in future usage. In addition to prior knowledge, comparing across different sources and platforms is another way of verifying information. Aligned with Choudhury et al.'s \cite{de2014seeking} findings that people co-use social media and search engines for health information, our study shows that people rely on their own trustful combination of information sources including certain websites, social media, and forum discussions. Building up one's own trustworthy combination of health information sources takes time and experience. ChatGPT as an emerging tool remains to be tested in people's daily health information searching tasks.       

ChatGPT is not a substitute for Google, but a stepping stone to more efficient information searching. The variance in trust score reveals a dynamic in people's trust towards different searching ways, but it does not necessarily isolate different means of searching for information. As people combine different resources for health information, it is meaningful to understand the position of ChatGPT in the big picture of combining multiple sources for health information, rather than viewing it as a single source. In our study, participants show a higher engagement with ChatGPT for understanding general information, but the accuracy of detailed information remains a concern. The same drawback has been pointed out by Jin' et al. \cite{jin2023retrieve} that ChatGPT contains inaccuracies for medical information, where they also suggested using ChatGPT as a summarization tool rather than for getting direct answers. Information-seeking is not a linear process. ChatGPT might be limited and not able to address the whole information process in the health domain; however, it would be meaningful to understand where ChatGPT can be situated toward providing better support for health information needs. 

Understanding scenarios of how people search for health information helps integrate ChatGPT to support health information needs. As revealed by our study, people search online mainly for several types of health information such as lifestyles, and accident treatments, which they consider as non-severe medical problems. With more people relying on online resources for health information, it is pivotal to understand the scenarios of health information seeking before introducing AI-powered tools. Although a plethora of domain-specific AI-powered ideas has been developed to solve specific problems such as AI in mental health\cite{d2020ai}, and Chatbot as an information portal during the COVID-19 pandemic \cite{xiao2023powering}, the daily scenarios of health information-seeking reserve potential for LLM-powered tools to help.

Overall, ChatGPT has its limitations, but also the potential to play a role in people's daily health information seeking. An ecological view considering where and when ChatGPT can help is a crucial step to better support people's needs for health information.


\section{Limitations and future work}
 
While our study provides valuable insights into the difference in trust in online health information from two distinct search agents, it is important to acknowledge its limitations for a more nuanced understanding of the results. 
First, our research focused solely on Google and ChatGPT as representatives of traditional search engines and LLM-based conversational search, respectively. While these are prominent examples, it is important to note that there are hybrid approaches emerging in the field, like those seen with Bing, combining web browsing and LLMs to offer an enhanced search experience. Nevertheless, our study provides a crucial first step toward unraveling insights into human trust and reliance on information provided by traditional and LLM-powered search agents (even if separately), paving the way for future research to explore emergent hybrid LLM search technologies and their implications on human trust.
Second, our study focuses on health-related information, which is an important yet specific domain. The differences in trust perceptions may vary when considering other types of information or contexts, and this specificity could limit the generalizability of our findings. 
Third, our study does not evaluate the accuracy of responses from ChatGPT. ChatGPT may make up answers (termed as AI hallucination phenomena \cite{Hallucin28:online}), which is one concern shared by our participants.
This invites future work into trust dynamics across other information-seeking contexts, to examine more closely the accuracy of LLM responses and their impact on trust. The web searching capability in the latest version of ChatGPT might be able to offset some of these concerns, though they remain fallible.
Fourth, our research design involved a lab study followed by semi-structured interviews. While this method allowed for in-depth insights, it may not fully capture the complexity and spontaneity of real-world information-seeking behaviors. While it would be ideal for participants to engage in naturalistic search behaviors based on their needs, such approaches raise significant privacy concerns. To mitigate this, our study design is purposefully structured to not only address privacy concerns but also offer a controlled environment to understand trust development in online health information seeking.
Lastly, the influence of varying demographics, like age and education level, was not deeply considered. Although these factors could potentially influence trust in online information, addressing these variables was beyond the scope of our current work. Nevertheless, it provides an opportunity for subsequent research to explore how different demographic groups perceive and trust online health information differently, thus broadening the understanding of online health information trust across diverse user groups.



Derived from the above, there are several avenues for future research that can build on our findings: Firstly, it would be beneficial to investigate the long-term impact of repeated interactions with LLM-powered agents on user trust and attitudes. Secondly, we can further explore to what extent users perceive biases or hallucinations from information returned by search engines and LLM-powered agents, and how such perceived bias affects human trust and attitudes toward the respective search agents. Thirdly, future research could explore how different individual demographics, such as age, education level, and cultural background, may affect trust perceptions, and how different trust perceptions differ across various contexts. This could provide valuable insights into how system developers and service providers can tailor their systems to meet the needs of diverse user groups and foster trust among different contexts. More importantly, with the continuous and rapid advancements in AI technology, it is vital to continue validating and verifying such findings relating to trust perceptions, and how these may change over time with newer AI advancements.





\section{Conclusion}

In this research, we explored the critical issue of trust towards health-related information obtained from two different search agents, namely, the traditional search engine Google and the emergent LLM-powered conversational agent ChatGPT. With online searches becoming indispensable tools for individuals seeking health information amid medical resource constraints, understanding how people perceive trust in this context is important. 
Our study provides pivotal insights into the factors influencing trust judgments when individuals interact with different search agents. Through a mixed-methods, within-subjects lab study, we observed significant variations in trust levels in online health information among participants. Notably, information sourced from ChatGPT gained higher trust compared to Google. The type of search task, however, did not significantly impact the participants' trust perceptions.
The determinants of trust were multifaceted, encompassing participants’ previous experiences, the style of information presentation, and the mode of interaction with search agents during the information-seeking process. These elements played a crucial role in shaping trust in the obtained health information. 
In conclusion, our study highlights important considerations and factors that are essential for enhancing the effective and reliable search for personal health information in the current era where generative AI plays a significant role. 
The findings from this study contribute to our understanding of how users navigate online health information with LLM-powered interface. While the eventual impact of these insights will be determined over time, we hope that our observations can support developers, researchers, and health informatics professionals in the era of sophisticated generative AI.


\newpage



\bibliographystyle{ACM-Reference-Format}
\bibliography{sample-base}


\end{document}